\documentclass[prd,aps,preprint,superscriptaddress,showkeys,showpacs,
 preprintnumbers,amsmath,amssymb,floatfix]{revtex4-1}
\usepackage{graphicx}
\usepackage{epsfig}
\usepackage{amsfonts}
\usepackage{dsfont}
\usepackage{amsmath}
\usepackage{amssymb}
\usepackage[sort&compress]{natbib}
\usepackage{dcolumn}
\usepackage{multirow}
\usepackage{bigstrut}
\usepackage{mathrsfs}
\usepackage{type1cm}
\usepackage{placeins}
\usepackage{color}
\usepackage{rotating}
\usepackage{slashed}
\usepackage{subfigure}
\usepackage{dcolumn}
\usepackage[mathscr]{euscript}
\newcommand{\Ll}{\mathcal{L}}

\DeclareMathOperator*{\SumInt}{%
\mathchoice%
  {\ooalign{$\displaystyle\sum$\cr\hidewidth$\displaystyle\int$\hidewidth\cr}}
  {\ooalign{\raisebox{.14\height}{\scalebox{.7}{$\textstyle\sum$}}\cr\hidewidth$\textstyle\int$\hidewidth\cr}}
  {\ooalign{\raisebox{.2\height}{\scalebox{.6}{$\scriptstyle\sum$}}\cr$\scriptstyle\int$\cr}}
  {\ooalign{\raisebox{.2\height}{\scalebox{.6}{$\scriptstyle\sum$}}\cr$\scriptstyle\int$\cr}}
}

\begin{document}

\preprint{\today}

\title{Cosmic-ray fermion decay through tau-antitau emission with Lorentz violation}

\author{C. A. Escobar}
\author{J. P. Noordmans}
\author{R. Potting}
\affiliation{CENTRA, Departamento de F\'isica, Universidade do Algarve, 8005-139 Faro, Portugal}

\begin{abstract}
We study CPT and Lorentz violation in the tau-lepton sector of the
Standard Model in the context of the Standard-Model Extension,
described by a coefficient which is thus far unbounded by experiment.
We show that any non-zero value of this coefficient implies that,
for sufficiently large energies,
standard-model fermions become unstable against decay due to the
emission of a pair of tau-antitau leptons. 
We calculate the induced fermion energy-loss rate and we deduce the
first limit on the Lorentz- and CPT-violating coefficient.
\end{abstract}

\maketitle

\section{Introduction}

Lorentz symmetry is a fundamental ingredient in both the Standard Model and
General Relativity. Nevertheless, various candidate theories of
quantum gravity suggest Lorentz violation (LV) may occur at the
Planck level \cite{qgmodels}.
Consequently, there have been extensive searches for Lorentz-violating signals.
Many of these studies use a general effective-field-theory framework
called the Standard-Model Extension (SME) \cite{ColladayKostelecky1,sme}.
It contains all LV gauge-invariant effective operators that can be built
from the conventional Standard-Model fields,
coupled to vector and tensor coefficients that parametrize the LV.
Within the SME, consistency requirements such as gauge invariance,
stability and causality can be \mbox{maintained \cite{Ralf}}.
The SME also contains all CPT-violating operators,
because in any local interacting quantum field theory
CPT violation (CPTV) implies LV \cite{Gre02}.
The SME allows us to quantify, in the most general possible way,
the exactness of Lorentz and CPT symmetry in the form of
observational constraints on the Lorentz-violating coefficients
(LVCs)~\cite{datatables}.
Such restrictions on LV and CPTV may be used as a guide
in finding the correct theory of quantum gravity.

It is generally safe to assume the LVCs, if nonzero, will be very small.
Consequently, they will generally produce very small effects in
certain physical quantities.
However, there are also cases in which LVCs allow certain processes
to take place that are normally forbidden.
In earlier work \cite{W-emission} we showed how a CPT and LV operator
in the sector of weak gauge bosons results in the possibility that at sufficiently high energy fermions that couple to these gauge bosons
can emit on-shell W bosons, resulting in their decay.
As such emissions should be able to occur for normally stable particles
such as the proton, this opens the possibility to bound the relevant SME
coefficient by considering the fact that protons are present in
Ultra-High-Energy Cosmic Rays (UHECR).

In this work we consider a similar decay process for an incoming (low-mass) fermion,
in which the latter is taken to emit a tau-antitau lepton pair
through an intermediate gauge boson (a photon or a $Z$ boson),
losing energy in the process.
Normally this process is, of course, forbidden in phase space.
For an elementary fermion (e.g.\ an electron) it can never occur
because of energy conservation.
For a hadron one might imagine it could occur if the emission
provokes the decay of the hadron into other particles.
Nevertheless, such a scenario is impossible for the proton,
because baryon number conservation imposes that in the final
decay products there has to be at least one (stable) baryon,
which has at least the mass of the proton.
In this paper we will assume that the propagation of the
gauge boson is described by its usual Lorentz-symmetric Lagrangian,
while the kinetic Lagrangian for the tau lepton is taken to include a CPT
and LV vector coefficient $b^\mu$ which is so far unbounded by experiment.
As we show, the presence of this coefficient will change
the dispersion relation of the tau and antitau leptons in such a way
that the fermion decay becomes possible in phase space above
a certain (large) threshold energy of the incoming fermion.
This threshold energy is inversely proportional to the size
of the components of the LV coefficient $b^\mu$,
which we will suppose to be very small.
In other words, we will assume the Earth-bound frame is concordant
\cite{concordant}.

A bound on $b^\mu$ can be obtained by applying this process to the proton.
The presence of protons in UHECRs implies their stability at
least up until the energy for which they are observed.
One crucial step is the demonstration that the decay rate
associated with this process is sufficiently large to exclude
the possibility that the proton would have survived the trip from
their astrophysical source to the Earth even when its momentum
exceeds the threshold for tau-antitau emission.

This paper is organized as follows.
In section \ref{sec:LVmodel} we analyze the kinetic Lagrangian
of the tau lepton in the presence of the Lorentz- and CPT-violating
$b^\mu$ coefficient and determine the dispersion relation
and the properties of the spin states. 
In section \ref{sec:kinematics} we analyze the kinematics of the
emission of a tau-antitau pair by an elementary fermion in the
presence of a $b^\mu$ coefficient in the kinetic term of the tau lepton.
In section \ref{sec:emission-rate} we evaluate the emission rate of
this process in two regimes of the momentum for an incoming
high-energy Dirac fermion.
This result is extended to the case of a proton
in section \ref{sec:proton}.
Finally, observational data from UHECRs are used to produce a bound
on the components of the $b^\mu$ coefficient.

\section{The Lorentz-violating model}
\label{sec:LVmodel}
We consider the following kinetic Lagrangian describing the Standard
Model tau lepton:
\begin{equation}
\mathcal{L}_{\rm LV} =
\bar\psi(i\slashed\partial+\slashed b\gamma^5-m_\tau)\psi\>,
\label{LVlagrangian}
\end{equation}
where the constant four-vector $b^\mu$ parametrizes a LV and
CPT-violating operator in the tau sector within the context
of the Standard-Model Extension~\cite{ColladayKostelecky1,sme}.
It can either be timelike, lightlike, or spacelike.

The energy-momentum tensor of the field $\psi$ is given by
\begin{equation}
T^{\mu\nu} = \frac{\partial \Ll}{\partial (\partial_\mu \psi)}\partial^\nu \psi - \eta^{\mu\nu}\Ll\>.
\end{equation}
Since $\partial_\mu T^{\mu\nu} = 0$ follows from Noether's theorem applied to invariance under spacetime translations, we have two conserved quantities: the Hamiltonian and the physical momentum:
\begin{align}
\label{hamilmom}
H &= \int d^3x\; T^{00} = \int d^3 x\; \mathcal{H} = \int d^3 x \bar{\psi}\left(i\vec{\gamma}\cdot \vec{\partial} + \gamma^5\slashed{b} + m_\tau \right) \psi\>, \\
P^i &= \int d^3x\; T^{0i} = \int d^3 x\; i\psi^\dagger\partial^i \psi = \int d^3x \; \pi \partial^i \psi\>,
\end{align}
where
\begin{equation}
\pi = \frac{\partial \Ll}{\partial (\partial_0 \psi)} = i\psi^\dagger\>,
\end{equation}
is the canonical momentum.  From the Euler-Lagrange equation we get the following Lorentz-violating equation of motion for $\psi$:
\begin{equation}
(i\gamma^\mu\partial_\mu - b_\mu \gamma_5\gamma^\mu- m_\tau)\psi = 0\>.
\label{eom}
\end{equation}
From this equation of motion, or from Eq.\ \eqref{hamilmom}, we derive that:
\begin{equation}
i\partial^0 \psi = \gamma^0(i\vec{\gamma}\cdot\vec{\partial} + m_\tau
+ \gamma^5\slashed{b})\psi \equiv \hat{H}[i\vec{\partial}]\psi\>.
\label{schrod}
\end{equation}
First, notice that $\hat{H}[i\vec{\partial}]$ is hermitian.
It therefore has real eigenvalues, i.e., all energies are real.

If we use the ansatz $\psi(x) = W(\vec{\lambda})e^{-i\lambda \cdot x}$,
the equation of motion for the spinors becomes
$(\slashed{\lambda} - \gamma^5\slashed{b} - m_\tau)
W^\alpha(\vec{\lambda}) = 0$ with $\alpha\in\{1,\ldots,4\}$,
or equivalently
\begin{equation}
\gamma^0(\slashed{\lambda} - \gamma^5\slashed{b} - m_\tau) W^\alpha(\vec{\lambda})
= (\lambda^0 - \hat{H}[\vec{\lambda}])W^\alpha(\vec{\lambda})
\equiv \Lambda_\alpha(\lambda)\, W^\alpha(\vec{\lambda})\>,
\label{offshelleig}
\end{equation}
with $\Lambda_\alpha(\lambda) = \lambda^0 - \omega_\alpha(\vec{\lambda})$,
where $\omega_\alpha(\vec{\lambda})$ are the roots of the dispersion relation.
This dispersion relation is given by
\begin{equation}
(\lambda^2-m_\tau^2-b^2)^2 - 4(\lambda\cdot b)^2 + 4\lambda^2 b^2 = \tilde\Lambda_+(\lambda)\tilde\Lambda_-(\lambda) = 0\>.
\label{dispersion_relation}
\end{equation}
Here $\tilde\Lambda_\pm(\lambda) = \lambda^2 - m_\tau^2 - b^2
\pm 2\sqrt{(\lambda\cdot b)^2 - b^2 \lambda^2}$,
where $\tilde\Lambda_+(\lambda)$ and $\tilde\Lambda_-(\lambda)$
each have two nondegenerate roots.
We will take the roots $\alpha=1,3$ to correspond to $\tilde\Lambda_+ = 0$,
and the roots $\alpha=2,4$ to $\tilde\Lambda_- = 0$.
(Below, we will see that $\alpha=1,2$ will correspond to particle states,
$\alpha=3,4$ to antiparticle states.)
In fact, it is not hard to see that
\begin{equation}
\prod_{\alpha=1}^4 \Lambda_\alpha = \tilde\Lambda_+ \tilde\Lambda_- \>.
\label{prod_eig}
\end{equation}
The $W^\alpha(\vec{\lambda})$ satisfy the off-shell eigenvalue equation \eqref{offshelleig}, as well as the on-shell equation:
\begin{equation}
\hat{H}[\vec{\lambda}]\,W^\alpha(\vec{\lambda}) =
\omega_\alpha(\vec{\lambda}) W^\alpha(\vec{\lambda}) \>.
\end{equation}
Since the operator on the left-hand side of Eq.\ \eqref{offshelleig} is hermitian,
it has four orthogonal eigenvectors $W^\alpha(\vec{\lambda})$:
\begin{equation}
W^\alpha(\vec{\lambda})^\dagger W^\beta(\vec{\lambda}) =
\delta^{\alpha\beta} N_\alpha(\vec{\lambda})\>,
\end{equation}
with $N_\alpha(\vec{\lambda})$ a real and positive normalization factor to be chosen later.
A straightforward analysis shows that this implies that
\begin{equation}
(\slashed{\lambda} - m_\tau - \gamma^5\slashed{b})^{-1} =
\frac{\operatorname{Adj}(\slashed{\lambda} - m_\tau - \gamma^5\slashed{b})}
{\det(\slashed{\lambda} - m_\tau - \gamma^5\slashed{b})} =
\sum_{\alpha=1}^4
\frac{W^\alpha(\vec{\lambda})\overline{W}^\alpha(\vec{\lambda})}
{N_\alpha(\vec{\lambda})\Lambda_\alpha(\lambda)}\>,
\end{equation}
with $\overline{W}^\alpha(\vec{\lambda}) = W^\alpha(\vec{\lambda})^\dagger\gamma^0$.
Since $\det(\slashed{\lambda} - m_\tau - \gamma^5\slashed{b}) = 
\prod_{\alpha=1}^4\Lambda_\alpha(\lambda)$, we find that 
\begin{equation}
 W^\alpha(\vec{\lambda})\overline{W}^\alpha(\vec{\lambda}) =
 \left. \frac{N_\alpha(\vec{\lambda})\operatorname{Adj}(\slashed{\lambda} - m_\tau - \gamma^5\slashed{b})}{\prod_{\beta\neq \alpha}\Lambda_\beta(\lambda)}
 \right\rfloor_{\lambda^0 = \omega_\alpha(\vec{\lambda})} = 
%\left. \frac{N_\alpha(\vec{\lambda})\operatorname{Adj}(\slashed{\lambda} - m_\tau - \gamma^5\slashed{b})}{{\rm Tr}\left(\operatorname{Adj}(\slashed{\lambda} - m_\tau - \gamma^5\slashed{b})\gamma^0\right)}
%\right|_{\lambda^0 = \omega_\alpha(\vec{\lambda})} =
\left.  \frac{N_\alpha(\vec{\lambda})\operatorname{Adj}(\slashed{\lambda} - m_\tau - \gamma^5\slashed{b})}{4(\lambda^2 - m_\tau^2 + b^2)\lambda^0 -8 (b\cdot\lambda)b^0} \right\rfloor_{\lambda^0 = \omega_\alpha(\vec{\lambda})}\>,
\label{spinorprojector}
\end{equation}
where the last identity follows from taking the derivative of both sides
of Eq.\ \eqref{prod_eig} with respect to $\lambda_0$.
One can also verify that
\begin{equation}
\operatorname{Adj}(\slashed{\lambda} - m_\tau - \gamma^5\slashed{b})
= (\slashed{\lambda} + m_\tau - \gamma^5 \slashed{b})
(\lambda^2 - m_\tau^2 - b^2 - 2i\gamma^5 \sigma^{\mu\nu} b_\mu \lambda_\nu)
\label{remarkable_identity}
\end{equation}
with $\sigma^{\mu\nu} = \frac{i}{2}[\gamma^\mu,\gamma^\nu]$.
Notice that it follows easily from the equation of motion that 
\begin{equation}
2i\gamma^5\sigma^{\mu\nu}b_\mu \lambda_\nu W^\alpha(\vec{\lambda})
= -(\lambda^2 - m_\tau^2 - b^2)W^\alpha(\vec{\lambda})\>.
\end{equation}
Therefore, the spinors $W^\alpha(\vec{\lambda})$ are eigenstates,
with eigenvalues $\pm 1$, of the CPT-violating operator
\begin{equation}
\mathcal{P} = \frac{i\gamma^5\sigma^{\mu\nu}b_\mu \lambda_\nu}{\sqrt{(\lambda\cdot b)^2 - \lambda^2 b^2}}\>,
\end{equation}
where all quantities are put on the relevant mass shell.
For purely timelike $b^\mu$, $\mathcal{P}$ can be seen to correspond
to $\operatorname{sgn}(b^0)$ times the helicity operator.
We take the spinors to satisfy
\begin{equation}
\mathcal{P}W_\alpha(\vec{\lambda}) = -(-1)^\alpha W_\alpha(\vec{\lambda})\>,
\qquad(\alpha=1,\ldots,4)\>.
\end{equation}
We now take the normalizations for the spinors to be
\begin{equation}
N^\alpha(\vec{\lambda}) = \left\{ \begin{array}{ccc} \displaystyle
\left|\left.\frac{\partial\tilde\Lambda_+(\lambda)}{\partial\lambda^0}\right\rfloor_{\lambda^0
= \omega_\alpha(\vec{\lambda})}\right|\  & \textrm{for }\alpha=1,3\>; \\[18pt]
\displaystyle\left|\left.\frac{\partial\tilde\Lambda_-(\lambda)}{\partial\lambda^0}\right\rfloor_{\lambda^0
= \omega_\alpha(\vec{\lambda})}\right|\  & \textrm{for }\alpha=2,4\>.
\end{array}\right.
\end{equation}
Since the denominator in Eq.\ \eqref{spinorprojector} is equal to
$\frac{\partial}{\partial \lambda^0}\left(\tilde\Lambda_+(\lambda)\tilde\Lambda_-(\lambda)\right)$,
we see that
\begin{equation}
 W^\alpha(\vec{\lambda})\overline{W}^\alpha(\vec{\lambda}) = \left\{\begin{array}{ccc}
\displaystyle\left. \frac{\operatorname{Adj}(\slashed{\lambda} - m_\tau - \gamma^5\slashed{b})}{2(\lambda^2 - m_\tau^2 - b^2)} \right\rfloor_{\lambda^0 = \omega_\alpha(\vec{\lambda})} & {\rm for} & \alpha=1,2\>; \\[12pt]
\displaystyle {}-\left. \frac{\operatorname{Adj}(\slashed{\lambda} - m_\tau - \gamma^5\slashed{b})}{2(\lambda^2 - m_\tau^2 - b^2)} \right\rfloor_{\lambda^0 = \omega_\alpha(\vec{\lambda})}\  & {\rm for} & \alpha=3,4\>.
\end{array}\right.
\label{spinorprojector2}
\end{equation}
By using Eq.\ \eqref{remarkable_identity} and the equation of motion
it is then easy to see that
\begin{equation}
\overline{W}^\alpha(\vec{\lambda})W^\alpha(\vec{\lambda}) =
\left\{\begin{array}{ccc}
2m_\tau\  & \textrm{for } \alpha=1,2\>; \\
-2m_\tau\  & \textrm{for } \alpha=3,4\>.
\end{array}\right.
\end{equation}

The dispersion relation \eqref{dispersion_relation} is invariant under $\lambda \rightarrow -\lambda$.
So for each positive solution $\lambda^0 = \omega(\vec{\lambda})$ there is
another negative solution $\lambda^0 = -\omega(-\vec{\lambda})$.
As usual we take these to correspond to particle and antiparticle energies. 
We label the particle energies by a $u$ and the antiparticle energies by
a $v$, while we differentiate between helicity states by a $\pm$ sign,
corresponding to the eigenvalue of the operator $\mathcal{P}$.
It follows that:
\begin{equation}
\begin{split}
E^u_+(\vec{p}) &= E^v_-(\vec{p}) = \omega_1(\vec{p},b) = -\omega_3(-\vec{p},b)\>, \\
E^u_-(\vec{p}) &= E^v_+(\vec{p}) = \omega_2(\vec{p},b) = -\omega_4(-\vec{p},b)\>.
\label{particle-energies}
\end{split}
\end{equation}
The corresponding redefined spinors are
$u_+(\vec{p}) = W^1(\vec{p})$,
$u_-(\vec{p}) = W^2(\vec{p})$,
$v_+(\vec{p}) = W^4(-\vec{p})$,
and $v_-(\vec{p}) = W^3(-\vec{p})$. 

Notice that the labels of the antiparticle states in \eqref{particle-energies}
have flipped sign because of the sign change under the identification
$\lambda^\mu \to -p^\mu$, in accordance with the fact that 
$\mathcal{P}$ flips sign when $\lambda^\mu$ does.
This obviates the fact that corresponding
particle and antiparticle states have different energies,
in accordance with the CPT-violating nature of the Lagrangian
$\mathcal{L}_{\rm LV}$.
Viewed this way, the equalities $E^u_\pm(\vec{p}) = E^v_\mp(\vec{p})$
in \eqref{particle-energies} could be considered rather coincidental
in the presence of CPT and Lorentz violation.
For instance, in the presence of an additional CPT-violating $a^\mu$
coefficient these equalities cease to hold \cite{ColladayKostelecky1}.

To the first no trivial order in the LV coefficient, the particle energies \eqref{particle-energies} can be easily
evaluated and, for arbitrary $b^\mu$, they are given by
\begin{equation}
E^{u}_\pm = E^{v}_\mp = \sqrt{\vec{\lambda}^2+m_\tau^2}\mp 
\frac{\sqrt{\bigg(b_0\sqrt{\vec{\lambda}^2+m_\tau^2}-\vec{\lambda}\cdot\vec{b}\bigg)^2-b^2m_\tau^2}}{\sqrt{\vec{\lambda}^2+m_\tau^2}}\>.
\label{E-general}
\end{equation}
Exact expressions for the energy can be written down for the cases
$b_0=0$, $\vec b = 0$ or $b^0=\pm |\vec b|$ (see \cite{ColladayKostelecky1},
where also explicit expressions for the corresponding eigen-spinors are derived),
but we will not need them in this work.

\section{Kinematics}
\label{sec:kinematics}

\begin{figure}%[h]
  \centering
    \includegraphics[width=0.5\textwidth]{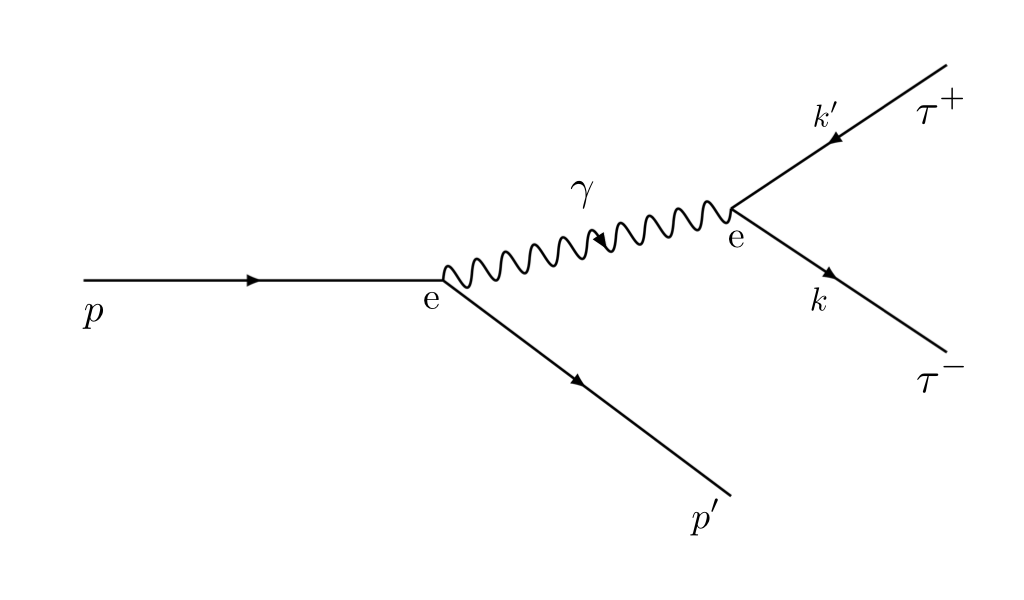}
    \vspace{-0.5cm}
  \caption{Feynman diagram for the photon-mediated emission process of a 
  $\tau^-,\tau^+$ pair with Lorentz-violating dispersion relation
  by a Dirac fermion of incoming momentum $p$ and outgoing momentum $p'$.}
  \label{fig:decay}
\end{figure}

We aim to calculate the rate at which a high-energy Dirac fermion
with mass $m$ emits a pair tau-antitau described by the Lagrangian \eqref{LVlagrangian}.
As a first step, we take the process to be mediated by a photon.
The corresponding Feynman diagram is indicated in figure~\ref{fig:decay}.
In the standard Lorentz-symmetric case,
this process is forbidden by energy-momentum conservation.
However, we will assume that the tau and anti-tau leptons in the final state
are described by the CPT and LV Lagrangian \eqref{LVlagrangian}.
This implies the modified dispersion relations defined by
Eqs.\ \eqref{particle-energies}.
In the following we will see that these imply the possibility of the emission process
to proceed at sufficiently high energy for the incoming fermion,
at least if we take the final spin states such that the tau and anti-tau
have spacelike four-momentum.

Let us denote the momenta of the incoming and outgoing fermion by $p$ and $p'$,
and those of the tau and anti-tau by $k$ and $k'$.
We parametrize the 3-momenta as follows:
\begin{equation}
\begin{split}
\vec{p} &= (0,0,p);\qquad \qquad \vec{p}\,'=(-(k_x+k_x'),-(k_y+k_y'),\lambda p)\>,  \\
\vec{k} &= (k_x,k_y,\alpha p);\qquad \qquad \vec{k}'=(k_x', k_y', \beta p)\>,
\label{parameterization-momenta}
\end{split}
\end{equation}
with $0<\alpha,\beta,\lambda<1$ and $\alpha+\beta+\lambda=1$. 

As we will be interested in the limit in which the incoming momentum
$p$ becomes ultra-relativistic,
we approximate the corresponding dispersion relations by 
\begin{equation}
\begin{split}
p_0 &\approx  p+\frac{m^2}{2p}\>, \\
p_0' &\approx \lambda p+\frac{(k_x+k_x')^2+(k_y+k_y')^2+m^2}{2\lambda p}\>,\\ 
{k_0}_\pm &\approx \alpha p+\frac{k_x^2+k_y^2+m_\tau^2}{2\alpha p} - (-1)^{\alpha_k}\xi_{b,k}\>,\\ 
{k_0'}_\pm &\approx \beta p+\frac{{k_x'}^2+{k_y'}^2+m_\tau^2}{2\beta p} - (-1)^{\alpha_{k'}} \xi_{b,k'}\>.
\end{split}
\label{disp-high}
\end{equation}
Here the parameter $\xi_{b,k}$ is of the order of the size
of the components of the LV coefficient $b^\mu$.
More precisely,
\begin{equation}
\xi_{b,k} = |b_0 - |\vec{b}|\cos(\theta_{kb})|\>,
\label{xi's}
\end{equation}
where $\theta_{kb}$ is the angle between the vectors $\vec k$ and $\vec b$.
A similar identity holds for $\xi_{b,k'}$.
We see from \eqref{xi's} that $\xi_{b,k}$ depends on $b^\mu$ and the
relative orientation of $\vec k$ and $\vec b$.
The parameters $\alpha_k$ and $\alpha_{k'}$ take the values 1 or 2,
indicating the helicity state of the tau and anti-tau (see section \ref{sec:LVmodel}).

The parametrization \eqref{parameterization-momenta} assures 3-momentum
conservation.
A supplementary condition comes from energy conservation $p_0 = p_0' + k_0 + k_0'$,
which yields, at lowest nontrivial order, the relation
\begin{align}
(-1)^{\alpha_k}\xi_{b,k}+(-1){^{\alpha_{k'}}} \xi_{b,k'} + \frac{m^2}{2p}
& = \frac{(k_x+k'_x)^2 + (k_y+k'_y)^2 + m^2}{2\lambda p}
\nonumber\\
&\qquad\qquad {} + \frac{k_x^2 + k_y^2 + m_\tau^2}{2\alpha p}
+ \frac{(k'_x)^2 + (k'_y)^2 + m_\tau^2}{2\beta p}\>.
\label{energy-conservation}
\end{align}
This condition can only be satisfied if the spin states are chosen
such that the first two terms are each chosen to be positive,
so that $(-1)^{\alpha_k}\xi_{b,k} = |\xi_{b,k}|$ and
$(-1){^{\alpha_{k'}}} \xi_{b,k'} = |\xi_{b,k'}|$. 
In other words, the tau-antitau emission can only take place when
the latter are each in particular helicity states.

Under this condition, we can now determine what is the minimum,
threshold value of the incoming momentum $p$ for which condition
\eqref{energy-conservation} can be satisfied.
It is not hard to verify that this happens when
\begin{equation}
k_x=k_x'=k_y=k_y'=0\>, \qquad \alpha=\beta=\frac{m_\tau}{m+2m_\tau}\>.
\label{threshold_values}
\end{equation}
Thus, at the threshold value the outgoing tau and antitau leptons 
move in exactly the same direction as the incoming fermion, and thus
all momenta are collinear. 
This in turn means
\begin{equation}
\theta_{kb}=\theta_{k'b}=\theta_{pb}=\theta_{p'b}
\qquad\textrm{(at threshold),}
\label{theta_threshold-values}
\end{equation}
and thus
\begin{equation}
\xi_{b,k}=\xi_{b,k'}\equiv\xi_{b,p}\qquad\textrm{(at threshold).}
\end{equation}
Using this we conclude from Eqs.\ \eqref{energy-conservation} and
\eqref{threshold_values} that
\begin{equation}
 p_{th} = \frac{m_\tau(m_\tau+m)}{|\xi_{b,p}|}\>.
 \label{p_th}
\end{equation}

It will be useful for the following to determine the energy loss
of the fermion with momentum $p>p_{th}$.
In order to do so, it is useful to go to the "brick-wall" observer
frame in which the four-momentum-transfer $q = k+k' = p-p'$ imparted
on the incoming fermion is purely space-like,
and thus no energy, only three-momentum is transferred.
That such a frame exists can be seen from the fact that both
the tau and anti-tau must have spacelike momentum for the emission
to be possible \cite{W-emission}.
In this frame the tau dispersion relation dips to negative energies
for a certain momentum range.

By analyzing the process in this frame and transforming back to the
concordant frame one finds that the incoming fermion suffers
an energy loss of at least
\begin{equation}
\Delta E_{min} = \frac{m_\tau^2}{|\xi_{b,p}|} \sim p_{th}\>.
\label{energyloss}
\end{equation}

Another interesting result that can be extracted is that the
three-momentum transfer in the brick-wall frame is of order
\begin{equation}
\frac{m_\tau^2}{\gamma|\xi_{b,p}|}\>,
\label{momentum-transfer}
\end{equation}
where $\gamma$ equals the gamma factor associated with the Lorentz 
transformation to the brick-wall frame.
At threshold the denominator in (\ref{momentum-transfer}) is of order $m_\tau$,
of the same order or larger than the fermion rest mass.
On the other hand, for ultra-high momenta $p\gg p_{th}$ the product
$\gamma|\xi_{b,p}|$ will generally be much larger than $m_\tau$, yielding
a very small (non-relativistic)  momentum transfer.
In this case the brick-wall frame corresponds practically to the fermion
rest frame.

\section{Emission rate}
\label{sec:emission-rate}

In order to be able to draw relevant physical conclusions about the
emission process, we need to estimate the rate of emission.
The  decay rate is given by
\begin{equation}
\Gamma =\frac{1}{2p_0}
\int \frac{d^3 k}{(2\pi)^3\Lambda'(k)}\frac{d^3 k'}{(2\pi)^3\Lambda'(k')}
\frac{d^3 p'}{(2\pi)^3 2p'_0}|\mathcal{M}|^2(2\pi)^4\delta^{(4)}(p-p'-k-k')\>,
\label{emission-rate}
\end{equation}
where the squared matrix element $|\mathcal{M}|^2$ is summed (averaged) over
the final (initial) fermion, but not over the final tau and anti-tau states.
The unconventional factors
\begin{equation}
\Lambda'(p) = \frac{\partial\Lambda_\pm(p)}{\partial p^0}\>,
\end{equation}
in the denominator define a (positive-definite) normalization
in which the phase space and the matrix element are separately observer
Lorentz invariant \cite{covquant}, i.e., invariant under simultaneous
Lorentz transformations of the momenta and the LV four-vector.
Explicit observer Lorentz covariance of the formalism will allow us to
transform to convenient observer frames later on.

The matrix element corresponding to the tree level process
mediated by the photon is given by
\begin{equation}
i \mathcal{M}=\bar u(p')(ie\gamma^\mu)u(p)
\frac{-i\eta_{\mu\nu}}{q^2}\bar u_\tau(k')(-ie\gamma^\nu)v_\tau(k)\>.
\label{matrix-element_photon}
\end{equation}
Here $u(p)$ and $u(p')$ are conventional Dirac spinors,
$q = k+k' = p-p'$ is the transferred momentum,
while $u_\tau(k')$ and $v_\tau(k)$ are described by the Lorentz and
CPT-violating kinetic Lagrangian \eqref{LVlagrangian}.
It follows from \eqref{matrix-element_photon} that the emission rate
\eqref{emission-rate} can be written as
\begin{equation}
\Gamma =\frac{1}{2p_0}\int\frac{d^3 k}{(2\pi)^3\Lambda'(k)}\frac{d^3 k'}{(2\pi)^3\Lambda'(k')}
\frac{8\pi e^4}{q^4}L^{\mu\nu}_\tau\,W_{\mu\nu}\>,
\label{emission-rate2}
\end{equation}
where 
\begin{equation}
L^{\mu\nu}_\tau = \operatorname{Tr}\left[\bar u_\tau(k')\gamma^\mu v_\tau(k)
\bar v_\tau(k)\gamma^\nu u_\tau(k')\right]
\label{L-tau}
\end{equation}
and 
\begin{equation}
W_{\mu\nu}=\frac{1}{8\pi}\int\frac{d^3 p'}{(2\pi)^3 2p'_0}
L^{\textrm{\tiny fermion}}_{\mu\nu} (2\pi)^4\delta^{(4)}(p - p' - k - k')
\label{W-munu}
\end{equation}
with
\begin{equation}
L^{\textrm{\tiny fermion}}_{\mu\nu} = \frac12 \sum_{\textrm{\tiny spins}}
\operatorname{Tr}\left[\bar u(p')\gamma_\mu u(p)\bar u(p)\gamma_\nu u(p')\right]\>.
\label{L-fermion}
\end{equation}

As it turns out, evaluating the decay rate for general values of
the incoming momentum is technically prohibitively complicated.
Fortunately, it is sufficient for our purposes to estimate the
order of magnitude of the rate.
We have been able to do evaluate the rate in two regimes:
a) when the incoming fermion momentum $p$ is just above threshold,
and b) in the asymptotic regime,
when the incoming fermion momentum is much larger than the threshold value.

We will first consider case a), corresponding to the conditions
\eqref{threshold_values} on the momenta of the particles.
Just above threshold, the matrix element squared $|\mathcal{M}|^2$ associated
with the Feynman diagram in Fig.\ \ref{fig:decay} does not vary strongly.
Thus, we can evaluate its value at threshold and obtain
\begin{equation}
\Gamma \approx \frac{e^4}{(2\pi)^5}\frac{1}{2p_0}
\bigg[\frac{1}{\Lambda'(k)}\frac{1}{\Lambda'(k')}\frac{1}{2p'_0}|\mathcal{M}|^2\bigg]
\bigg|_{th}\int d^3k d^3 k' d^3 p'\delta^{(4)}(p-p'-k-k')\>,
\end{equation}
where $th$ is defined by the conditions in Eqs.\ \eqref{threshold_values}.
By using the fact that the tau and antitau satisfy the
modified dispersion relations \eqref{disp-high},
as well as the following identity for the spinor bilinears
\begin{equation}
u_\tau^i(k) \bar{u}_\tau^i(k)=\frac{(\slashed{k}+m-\gamma^5\slashed{b})(k^2-m^2-b^2-2i\gamma^5\sigma^{\mu\nu}b_\mu k_\nu)}{2(k^2-m^2-b^2)}\>,
\label{bilinear-u}
\end{equation}
and
\begin{equation}
v_\tau^i(k) \bar{v}_\tau^i(k)=\frac{(\slashed{k}-m+\gamma^5\slashed{b})(k^2-m^2-b^2+2i\gamma^5\sigma^{\mu\nu}b_\mu k_\nu)}{2(k^2-m^2-b^2)}\>,
\label{bilinear-v}
\end{equation}  
which follow from Eqs.\ \eqref{spinorprojector2} and
\eqref{remarkable_identity}, one can derive that 
\begin{equation}
\bigg[\frac{1}{\Lambda'(k)}\frac{1}{\Lambda'(k')}\frac{1}{2p'_0}|\mathcal{M}|^2\bigg]\bigg|_{th}=\frac{2}{p^3}\bigg(\frac{m_\tau}{m}\bigg)^2\frac{1}{(\alpha\beta)^2}\bigg|_{th}=\frac{2}{p^3}\frac{(m+2m_\tau)^4}{m^2m_\tau^2}+ O\bigg(\frac{\xi^2}{p^3 m^2},\frac{m}{p^4}\bigg)\>.
\end{equation}
Moreover, one can show that
\begin{equation}
\int d^3k d^3 k' d^3 p'\delta^{(4)}(p-p'-k-k')
= p^3\frac{m^{11/2}m_\tau^3}{(m+2m_\tau)^{13/2}}\pi^3 R^4\>,
\end{equation}
where
\begin{equation}
R^2 = \mp\frac{4|\xi_{b,p}| p}{m^2}-\frac{4m_\tau(m+m_\tau)}{m^2}\>,
\end{equation}
so that
\begin{equation}
\Gamma =\frac{2\pi^3 e^4}{(2\pi)^5}	\frac{m^{\frac{7}{2}}m_\tau}{(m+m_\tau)^{\frac{5}{2}}} \frac{R^4}{p}\>.
\end{equation}
Taking $p = a\,p_{th}$, it then follows that 
\begin{equation}
\Gamma\approx \frac{e^4|\xi_{b,p}|}{\pi^2}\frac{m_\tau^2}{\sqrt{m(m+m_\tau)^3}} (a-1)^2 \theta(a-1) \>.
\label{rate-threshold}
\end{equation}

Now let us consider case b) for which the scale of the incoming
fermion momentum $p$ is much larger than the threshold momentum for
tau-antitau emission, or, equivalently,
that $|\xi_{b,p}|\gg m_\tau(m_\tau + m)/p$.
In this  case, we see from the energy conservation relation
\eqref{energy-conservation} that we can expect that the transverse momenta
of the outgoing particles should scale as $\sqrt{|\xi_{b,p}| p}$.
For this reason we introduce the rescaled,
dimensionless transverse momentum variables
\begin{equation}
\tilde k_{x,y} = \frac{k_{x,y}}{\sqrt{|\xi_{b,p}| p}}\>, \qquad\qquad
\tilde k'_{x,y} = \frac{k'_{x,y}}{\sqrt{|\xi_{b,p}| p}}\>.
\label{rescaled}
\end{equation}
Note that the transverse momentum components of the outgoing fermion can be
expressed in terms of $\tilde k_{x,y}$ and $\tilde k'_{x,y}$ by three-momentum
conservation.

In order to perform the calculation of the emission rate below,
we now make the approximation that the identities \eqref{theta_threshold-values}
hold approximately in the asymptotic ($p\gg p_{th}$) limit.
This will be the case because the transverse ($x,y$) components of $\vec k$
and $\vec k'$ scale as the square root of the incoming momentum $p$,
while the parallel ($z$) component is proportional to $p$.
Thus, the tau and anti-tau are emitted in a small forward cone along the
incoming momentum.
This means that we can take $\xi_{b,k} \approx \xi_{b,k'} \approx \xi_{b,p}$
as fixed quantities in the dispersion relation.

Taking now the high-$p$ limit, maintaining terms at lowest order in
$1/p$, it follows that
\begin{equation}
|\mathcal{M}|^2 = \frac{16 m_\tau^2 e^4(1-\alpha-\beta)}{\alpha\beta}
\frac{|\xi_{b,p}| p\, K (1+(1-\alpha-\beta)^2) + m^2(\alpha+\beta)^4}{(|\xi_{b,p}| p\, K + m^2(\alpha+\beta)^2)^2}\>,
\quad\quad\quad 
\end{equation}
where we have introduced
\begin{equation}
K = (\tilde k_x + \tilde k'_x)^2 + (\tilde k_y + \tilde k'_y)^2\>.
\end{equation}
The expression for the asymptotic decay rate becomes 
\begin{align}
\Gamma&=\frac{e^4m_\tau^2|\xi_{b,p}|^2 p^3}{(2\pi)^5}
\int\frac{d\alpha\,d\beta\,d\tilde k_x\,d\tilde k_y\,d\tilde k'_x\,d\tilde k'_y}{\alpha^2\beta^2}\nonumber\\
&\qquad\qquad\qquad\quad{}\times\frac{|\xi_{b,p}|p\, K(1+(1-\alpha-\beta)^2) + m^2(\alpha+\beta)^4}
{(|\xi_{b,p}| p\, K + m^2(\alpha+\beta)^2)^2}\delta(p-p^0-k^0-{k'}^0)\>.
\end{align}
The integral over the transverse momenta can now be performed,
and the resulting expression can be cast into the form 
\begin{align}
\Gamma &= \frac{2e^4m_\tau^2}{(2\pi)^3 p}
\int^{a_2^{-1}-1}_{a_1}dz\left(\frac{z}{1+z}\right)
\left[-\frac{\ln\left(a_2(1+z)\right)}{2} +
\ln\left(\sqrt{1-\frac{a_1}{z}}+\sqrt{1-\frac{a_1}{z}-a_2(1+z)}\right)\right]\nonumber \\
&\qquad\qquad\qquad{}\times\left[-1+\frac{1}{2}\left(\frac{(z+1)^2+z^2}{z(1+z)}\right)
\ln\left(\frac{z}{a_1}\right)\right]+\mathcal{O}\left(\frac{1}{p}\right)\>,
\label{rate-asymp-integral}
\end{align}
where we introduced the parameters
\begin{equation}
a_1=\frac{m^2}{4|\xi_{b,p}| p}\>,\qquad a_2=\frac{m_\tau^2}{|\xi_{b,p}| p}\>,
\end{equation}
which tend to zero in the large-$p$ limit.
For the case in which $m$ and $m_\tau$ are of the same order,
both $a_1$ and $a_2$ are of order $p_{th}/p$.

While exact analytic evaluation of the $z$-integral in
\eqref{rate-asymp-integral} is not feasible,
one can deduce that the dominant contribution is of the form
\begin{equation}
\frac{-\ln(a_2)}{a_2}\bigl(C_1+C_2\ln(a_1 a_2)\bigr)\>,
\label{log-formula}
\end{equation}
for some dimensionless constants $C_1$ and $C_2$
(here we excluded the prefactors in \eqref{rate-asymp-integral}).
Numerical fitting of the $z$-integral in \eqref{rate-asymp-integral}
to the formula \eqref{log-formula} confirms that an excellent fit
can be obtained.
\begin{table}[th]
\setlength{\tabcolsep}{12pt}
\begin{center}
  \begin{tabular}{ | c | c | c | }
    \hline
    \multicolumn{1}{|c|}{$a_1/a_2$} & \multicolumn{1}{c|}{$C_1$} & \multicolumn{1}{c|}{$C_2$} \\ \hline
    $0.025$ & $2.5211$ & $0.00759$ \\
    $0.138875$ & $1.9721$ & $-0.00060$ \\
    $0.25$ & $1.7969$ & $-0.00341$ \\
    $1.00$ & $1.4097$ & $-0.01004$ \\
    $2.5$ & $1.1738$ & $-0.01442$ \\
    \hline
  \end{tabular}
\end{center}
\caption{Fitted values of the parameters $C_1$ and $C_2$ in formula
\eqref{log-formula} as a function of the ratio $a_1/a_2$.
The value $a_1/a_2=0.138875$ corresponds to taking for $m$ the proton mass.
The value $a_1/a_2=2.5$ corresponds to taking for $m$ ten times the tau mass.}
\label{integral_table}
\end{table}
In Table \ref{integral_table} we list fitted values for
$C_1$ and $C_2$ as a function of the ratio $a_1/a_2 = m^2/(4m_\tau^2)$.
In obtaining the fit, we took $a=p/p_{th}$ distributed logarithmically
along the range $10^2<a<10^{10}$.
It follows that
\begin{equation}
\Gamma \sim \frac{2e^4|\xi_{b,p}|}{(2\pi)^3}
\bigl(-\ln(a_2)\bigr)\bigl(C_1+C_2\ln(a_1 a_2)\bigr)\>.
\label{rate-asymp}
\end{equation}
We see that the asymptotic form of the rate is proportional to $|\xi_{b,p}|$
times a quadratic expression in terms of $\ln(a)$. 

In conclusion, a charged fermion will start emitting tau-antitau pairs
if its momentum $p$ exceeds a certain threshold value $p_{th}$.
The emission rate is of the form
\begin{equation}
\Gamma = \frac{e^4 |\xi_{b,p}|}{2\pi^2} G(a)\theta(a-1)
= 8\alpha^2 |\xi_{b,p}| G(a)\theta(a-1)\>,
\end{equation}
(with $\alpha$ the fine-structure constant)
for a function $G(a)$ that satisfies
\begin{equation}
G(a) \approx \left\{\begin{array}{ll}
        \dfrac{2m_\tau^2}{\sqrt{m(m+m_\tau)^3}}(a-1)^2, & \textrm{if } a-1 \ll 1\>; \\[18pt]
        \dfrac{1}{2\pi}\bigl(-\ln(a_2)\bigr)
        \bigl(C_1+C_2\ln(a_1 a_2)\bigr),\ \ \ \  & \textrm{if } a\gg 1\>. 
        \end{array}\right.
\label{G}
\end{equation}

From the expressions \eqref{rate-threshold} and \eqref{rate-asymp}
we find that if $\mathcal{O}(|\xi_{b,p}|)\sim 10^{-10}\,$GeV,
corresponding to the bound we will find later on,
the typical decay time is 
\begin{equation}
t_p\sim10^{-11}\,\textrm{s}\>.
\label{decay-time}
\end{equation}
Moreover, it was shown at the end of the previous section that the emission
process implies an energy loss of at least
$m_\tau^2/(2|\xi_{b,p}|)\sim p_{th}$.
It follows that for such values of $|\xi_{b,p}|$
all fermions in a decay cascade will fall below threshold within
at most $a\times 10^{-11}\,$s.

\section{The proton}
\label{sec:proton}

The analysis in the previous sections applies to an incoming
fundamental charged Dirac fermion.
In fact, of more practical interest to us will be the case of an incoming
composite particle such as the proton.
In applying the formalism developed in the previous sections to an incoming
proton, a few adaptations are necessary.

First of all, the tensor $W_{\mu\nu}$ in expression \eqref{W-munu}
now applies to the photon coupling to the proton and becomes
\begin{equation}
W^{\mu\nu} = \frac{1}{8\pi}\sum_\sigma\SumInt_X
\langle {\rm p}(p,\sigma)|J^\nu(-q)|X\rangle
\langle X|J^\mu(q)|{\rm p}(p,\sigma)\rangle\>,
\label{W-munu_hadron}
\end{equation}
where $|{\rm p}(p,\sigma)\rangle$ is a proton state with momentum $p$
and spin $\sigma$, $J^\mu(p)$ is the hadronic current,
and $\SumInt_X$ represents a sum over the possible hadronic final states $X$
along with the corresponding integrations over phase space.

Insight into the final hadronic state can be gained from the discussion
at the end of section \ref{sec:kinematics}.
As it was argued there, for incoming momenta far above the threshold $p_{th}$
(that is, for $a\gg 1$),
the momentum transfer $\vec k + \vec k'$ in the rest frame of the fermion
will be tiny.
This means the impact of the tau-antitau emission on the proton will be small
so that it will remain intact.
That is, the final state $|X\rangle$ corresponds to a (ground-state) proton.
The applicable hadron current $J^\mu$ then can be written with the
usual proton structure functions:
\begin{equation}
J^\mu(q)= e\,\bar u(p')\left[F_1\bigl(q^2\bigr)\gamma^\mu
+ \frac{\kappa}{2m}F_2\bigl(q^2\bigr)i\sigma^{\mu\nu}q_\nu\right] u(p)
\label{proton-current}
\end{equation}
where the proton anomalous magnetic moment $\kappa = 1.79$.
Also, as $q$ will be tiny, we can take the limits $F_1(0) = F_2(0) = 1$.
Thus, the only change with respect to the calculation of the rate
as compared to the case of a fundamental Dirac fermion will be the
$F_2$ term.
However, as the latter is proportional to the momentum transfer components
$q_\nu$,
it can be expected to be very small compared to the first term,
and can be safely ignored.
We conclude that for ultra-high proton momenta $p\gg p_{th}$,
the emission rate is given by expression \eqref{emission-rate} and
that the energy of the proton will cascade down as described at the end
of section \ref{sec:emission-rate}.

As we saw in section \ref{sec:kinematics}, for momenta close to the
threshold for emission, the momentum transfer in
the brick-wall frame will become of the order of the proton rest mass.
As a consequence, it is more adequate to consider the tau-antitau emission
as a (deeply) inelastic process, resulting in the breakup of the proton.
In this case we can use the parton model to evaluate $W^{\mu\nu}$ in
expression \eqref{W-munu_hadron}.
This essentially involves calculating the decay rate of an elementary quark
that carries a fraction $x$ of the longitudinal proton momentum.
We can thus use many of the results obtained for the elementary fermion rate.
For a pedagogical introduction to parton-model calculations,
we refer to Ref.~\cite{PeskinSchroeder}.

The resulting emission rate in the context of the parton model becomes
\begin{equation}
\Gamma = \frac{e^4|\xi_{b,p}|}{2\pi^2}
\sum_q\int_0^1 dx\bigl(f_q(x)+\bar f_q(x)\bigr)
\tilde G_q^{ee}(ax)\theta(ax-1)\>.
\label{rate_proton}
\end{equation}
Here the functions $f_q(x)$ and $\bar f_q(x)$ are the parton
distributions functions (PDFs) for the quarks and antiquarks
of flavor $q$, respectively, representing the probability of
finding a quark with momentum fraction $x$ inside the proton.
The PDFs are assumed to be independent of $p^2$,
which is a good approximation to leading order in the strong
coupling constant.
The function $G_q(ax)$ corresponds to the absolute-value
squared of the first term in \eqref{matrix-element_hadron},
and is equal to the function \eqref{G} with the 
substitution $m\to x m$, and an extra multiplication by the square
of the quark charge fraction.

As the integral over $x$ in Eq.\ \eqref{rate_proton} runs up to 1,
the maximum momentum of the parton involved in the emission of the
tau-antitau pair equals that of the incoming proton itself.
As a consequence the threshold proton momentum for tau-antitau
emission will be given by formula \eqref{p_th},
where $m$ is to be identified with the proton mass.
At such values of $x$ close to 1,
the proton PDFs for valence quarks decay to zero approximately
as a constant times $(1-x)^{c_q}$ with $c_u \approx 4$ and
$c_d \approx 5$ \cite{pdf}.
As a consequence, the integral over $x$ in Eq.\ \eqref{rate_proton}
yields decay rates just above threshold that are suppressed
as compared to the situation in which the proton would have
been an elementary particle.

\begin{figure}%[h]
  \centering
    \includegraphics[width=0.65\textwidth]{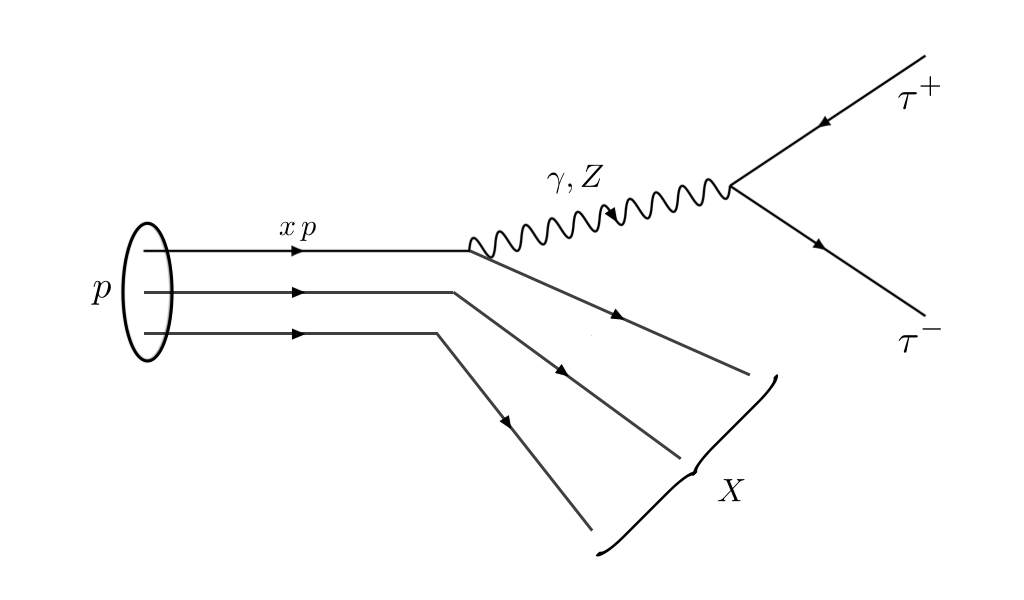}
    \vspace{-0.5cm}
  \caption{Feynman diagram for the emission process of a tau-antitau
  pair by a proton of incoming momentum $p$,
  mediated by a photon or a $Z$ boson.}
  \label{fig:proton-decay}
\end{figure}
Finally, we note that in the above we only considered the emission process
in which the tau-antitau emission is mediated by a virtual photon.
In fact, there is the additional possibility that the emission takes
place with an intermediate $Z$ boson instead (see Fig.\ \ref{fig:proton-decay}).
This means that the transition amplitude $\mathcal{M}$ is given by the sum
of two terms:
\begin{align}
i \mathcal{M}&=\langle X|ieJ^\mu_{EM}(q)|p(p,\sigma)\rangle
\frac{-i\eta_{\mu\nu}}{q^2}\bar u_\tau(k')(-ie\gamma^\nu)v_\tau(k)
\nonumber\\
&\quad{}+
\langle X|igJ^\mu_{Z}(q)|p(q,\sigma)\rangle
\frac{-i\eta_{\mu\nu}}{q^2-m_Z^2}
\bar u_\tau(k')(-i\frac{g}{2}\gamma^\nu)\left(c_V^\tau-c_A^\tau\gamma^5\right)v_\tau(k)\>.
\label{matrix-element_hadron}
\end{align}
Here $J^\mu_{EM}$ corresponds to the electromagnetic current considered
in \eqref{W-munu_hadron} and in the previous sections,
while $J^\mu_{Z}$ applies to the current coupling to the $Z$ boson.
In the context of the parton model, the emission rate will now be of the form
\begin{equation}
\Gamma = \frac{|\xi_{b,p}|}{2\pi^2}
\sum_q\int_0^1 dx\bigl(f_q(x)+\bar f_q(x)\bigr)
\bigl(e^4\tilde G_q^{ee}(ax) + g^4\tilde G_q^{gg}(ax)
+ e^2g^2\tilde G_q^{eg}(ax)\bigr)\theta(ax-1)\>,
\label{rate_proton_gamma+Z}
\end{equation}
where the function $\tilde G_q^{ee}(ax)$ is the same as $G_q(ax)$ in
Eq.\ \eqref{rate_proton}, while
the terms with $\tilde G_q^{gg}$ and $\tilde G_q^{eg}$ correspond to
the absolute-value squared of the second term in \eqref{matrix-element_hadron},
and a cross (interference) term, respectively.
They can be expected to be of the same general form as $G_q^{ee}(ax)$.
For our purposes it is not necessary to compute them explicitly,
as we will only need a lower bound on the emission rate.
It is sufficient to note that the additional terms are independent expressions
proportional to the independent quantities $g^4$ and $e^2g^2$,
and thus a cancellation would imply an extreme finetuning between
$G_q^{ee}$, $G_q^{gg}$ and $G_q^{eg}$ (the latter would have to be negative).

In fact, note that in the asymptotic limit $p\gg p_{th}$ the momentum
transfer $q^\mu$ in the brick-wall frame is very small.
This means that the $Z$-boson propagator will be very suppressed compared
to the photon propagator,
which implies in turn that the decay rate will be dominated by the
photon exchange diagram for $p\gg p_{th}$.

\section{Observational limits from ultra-high-energy cosmic rays}
\label{observational-limits}

We can obtain a bound on $|\xi_{b,p}|$ by observing that a cosmic-ray proton
that has an energy below threshold has zero probability
to disintegrate by tau-antitau emission,
and can thus reach Earth unimpeded.
However, with an energy above threshold,
the proton can emit tau-antitau pairs,
possibly disintegrating eventually,
until all protons in the decay cascade have fallen below threshold .
This means that it cannot reach Earth if its mean free path
is much smaller than the distance $D$ from its source to Earth.

Since many ultra-high-energy cosmic-ray (UHECR) particles with energies
above $57\,\mathrm{EeV}\equiv |\vec{p}|_{obs}$ have been observed,
coming more or less from all directions \cite{uhecrdetected},
we can take it as a first estimate for the lower bound for $E_{\mathrm{th}}$.
It follows that
\begin{equation}
|\xi_{b,p}| < \frac{m_\tau(m_\tau + m)}{|\vec{p}|_{obs}}
\approx 8.5 \times 10^{-11}\,\mathrm{GeV}\>.
\label{kappabound}
\end{equation}
This bound can only be relaxed if the mean free path of protons above
threshold is not much smaller than $D$.
From Eq.~\eqref{rate_proton} we see
that the mean lifetime of protons (in Earth's frame) $t_p$
is still proportional to $|\xi_{b,p}|^{-1}$, but is enhanced,
mainly by the minute values of the PDFs at large $x$.
A very conservative estimation that comes from comparing to the
elementary-fermion decay time \eqref{decay-time}
(involving only photon exchange) gives a mean free path of
\begin{equation}
L \simeq c \,t_p \times 10^{15} \approx 3\times10^{9}\,\textrm{km}
\label{L}
\end{equation}
which is of the order of the size of the solar system.
Clearly, in such a scenario, protons with an energy above threshold
will not be able to reach Earth from any viable UHECR source.
We thus obtain a bound on all four components of the LVC:
\begin{equation}
|b^\mu| < 8.5 \times 10^{-11}\,\textrm{GeV}\>.
\label{limitresult}
\end{equation}
We note that we have to assume that at least some of the
detected UHECRs are protons and that these have a sufficient
spread in arrival direction.
Although the mass content of the UHECRs, in particular at high energies, 
is still largely unexplored \cite{uhecrmass},
it seems very unlikely that such a significant low-mass
component is completely absent.
Moreover, even if this were the case, one can calculate a
decay rate equivalent to Eq.~\eqref{rate_proton},
by using nuclear PDFs \cite{npdf}.
Since $L \ll D$ by many orders of magnitude,
none of this will change our result in Eq.~\eqref{limitresult}.

\section{Conclusion}
\label{sec:conclusion}

In this paper, we investigated Lorentz and CPT-violation in the 
electroweak sector.
In particular, we focussed on a hitherto unbounded Lorentz-violating
coefficient in the tau sector of the Standard-Model Extension.
By considering the emission of tau-antitau pairs by high-energy particles
in the context of  UHECR observations leads to a new bound
on the coupling $b^\mu$ in the tau sector of about $10^{-10}\,$GeV.
Our method relies on the fact that any hypothetical Lorentz violation in
the tau sector parametrized by the $b^\mu$ coefficient turns a process that
is normally forbidden in phase space possible for sufficiently high momentum.
It is worth pointing out that formula \eqref{kappabound} can be applied
to the muon and the electron sectors as well,
under the substitutions $m_\tau \to m_\mu$ and $m_\tau \to m_e$.
This leads to the noncompetitive bounds $1.9\times10^{-12}\,$GeV and
$8.4\times10^{-15}\,$GeV for the components of the $b^\mu$ parameter in
the muon sector and the electron sector, respectively.
However, we do expect that it should be possible to derive new
and/or competitive bounds on other SME coefficients,
either in the tau sector or for other particles, by applying this method.

\acknowledgments
This work is supported in part by the Funda\c c\~ao para a Ci\^encia e
a Tecnologia of Portugal (FCT) through projects UID/FIS/00099/2013 and
SFRH/BPD/101403/2014 and program POPH/FSE.
C.\ A.\ E.\ was supported by a CONACyT postdoctoral grant No.\ 234745.


\begin{thebibliography}{99}
  
\bibitem{qgmodels}
  V.~A.~Kosteleck\'y and S.~Samuel,
  %``Spontaneous Breaking of Lorentz Symmetry in String Theory,''
  Phys.\ Rev.\ D {\bf 39}, 683 (1989);
  %%CITATION = doi:10.1103/PhysRevD.39.683;%%
  V.~A.~Kostelecky and R.~Potting,
  %``CPT and strings,''
  Nucl.\ Phys.\ B {\bf 359}, 545 (1991).
  %%CITATION = doi:10.1016/0550-3213(91)90071-5;%%
  J.~R.~Ellis, N.~E.~Mavromatos, and D.~V.~Nanopoulos,
  %``Search for quantum gravity,''
  Gen.\ Rel.\ Grav.\  {\bf 31}, 1257 (1999);
  %%CITATION = doi:10.1023/A:1026720723556;%%
  R.~Gambini and J.~Pullin,
  %``Nonstandard optics from quantum space-time,''
  Phys.\ Rev.\ D {\bf 59}, 124021 (1999);
  %%CITATION = doi:10.1103/PhysRevD.59.124021;%%
  C.~P.~Burgess, J.~M.~Cline, E.~Filotas, J.~Matias, and G.~D.~Moore,
  %``Loop generated bounds on changes to the graviton dispersion relation,''
  JHEP {\bf 0203}, 043 (2002).
  %%CITATION = doi:10.1088/1126-6708/2002/03/043;%%

\bibitem{ColladayKostelecky1}
  D.\ Colladay and V.~A.\ Kosteleck\'y,
  Phys.\ Rev.\ D {\bf 55}, 6760 (1997).
   %%CITATION = doi:10.1103/PhysRevD.55.6760;%%
 \bibitem{sme} 
   D.\ Colladay and V.~A.\ Kosteleck\'y,
  Phys.\ Rev.\ D {\bf 58}, 116002 (1998).
 %%CITATION = doi:10.1103/PhysRevD.58.116002;%%
 \bibitem{Ralf} 
  V.~A.\ Kosteleck\'y and R. Lehnert,
  Phys.\ Rev.\ D {\bf 63}, 065008 (2001).


\bibitem{Gre02}
  O.~W.~Greenberg,
  %``CPT violation implies violation of Lorentz invariance,''
  Phys.\ Rev.\ Lett.\  {\bf 89}, 231602 (2002).
  %%CITATION = doi:10.1103/PhysRevLett.89.231602;%%

\bibitem{datatables}
  V.~A.~Kostelecky and N.~Russell,
  %``Data Tables for Lorentz and CPT Violation,''
  Rev.\ Mod.\ Phys.\  {\bf 83}, 11 (2011)
  [2018 edition: arXiv:0801.0287v11 [hep-ph]].
  %%CITATION = doi:10.1103/RevModPhys.83.11;%%

\bibitem{W-emission} 
  D.~Colladay, J.~P.~Noordmans and R.~Potting,
  %``Cosmic-ray fermion decay by emission of on-shell W bosons with CPT violation,''
  Phys.\ Rev.\ D {\bf 96}, 035034 (2017).
  % doi:10.1103/PhysRevD.96.035034
  % [arXiv:1704.03335 [hep-ph]].
  %%CITATION = doi:10.1103/PhysRevD.96.035034;%%
  
\bibitem{concordant} 
  V.~A.~Kostelecky and R.~Lehnert,
  Phys.\ Rev.\ D  {\bf 63}, 065008 (2001).

\bibitem{covquant}
  D.~Colladay, P.~McDonald, J.~P.~Noordmans, and R.~Potting,
  %``Covariant Quantization of CPT-violating Photons,''
  Phys.\ Rev.\ D {\bf 95}, 025025 (2017).
  %%CITATION = doi:10.1103/PhysRevD.95.025025;%%

\bibitem{PeskinSchroeder} 
  M.~E.~Peskin and D.~V.~Schroeder,
  {\it An Introduction to quantum field theory\/}, (Addison-Wesley, 1995).

\bibitem{pdf}
  D.~Stump, J.~Huston, J.~Pumplin, W.~K.~Tung, H.~L.~Lai, S.~Kuhlmann, and J.~F.~Owens,
  %``Inclusive jet production, parton distributions, and the search for new physics,''
  JHEP {\bf 0310}, 046 (2003);
  %%CITATION = doi:10.1088/1126-6708/2003/10/046;%%
  S.~Alekhin, K.~Melnikov, and F.~Petriello,
  %``Fixed target Drell-Yan data and NNLO QCD fits of parton distribution functions,''
  Phys.\ Rev.\ D {\bf 74}, 054033 (2006);
  %%CITATION = doi:10.1103/PhysRevD.74.054033;%%
  J.~F.~Owens, J.~Huston, C.~E.~Keppel, S.~Kuhlmann, J.~G.~Morfin, F.~Olness, J.~Pumplin, and D.~Stump,
  %``The Impact of new neutrino DIS and Drell-Yan data on large-x parton distributions,''
  Phys.\ Rev.\ D {\bf 75}, 054030 (2007).
  %%CITATION = doi:10.1103/PhysRevD.75.054030;%%

\bibitem{uhecrdetected} 
  A.~Aab {\it et al.} [Pierre Auger Collaboration],
  %``Searches for Anisotropies in the Arrival Directions of the Highest Energy Cosmic Rays Detected by the Pierre Auger Observatory,''
  Astrophys.\ J.\  {\bf 804}, 15 (2015);
  %%CITATION = doi:10.1088/0004-637X/804/1/15;%%
  T.~Abu-Zayyad {\it et al.} [Telescope Array Collaboration],
  %``Correlations of the Arrival Directions of Ultra-high Energy Cosmic Rays with Extragalactic Objects as Observed by the Telescope Array Experiment,''
  Astrophys.\ J.\  {\bf 777}, 88 (2013).
  %%CITATION = doi:10.1088/0004-637X/777/2/88;%%

\bibitem{uhecrmass} 
  A.~Aab {\it et al.} [Pierre Auger Collaboration],
  %``Depth of maximum of air-shower profiles at the Pierre Auger Observatory. II. Composition implications,''
  Phys.\ Rev.\ D {\bf 90}, 122006 (2014);
  %%CITATION = doi:10.1103/PhysRevD.90.122006;%%
  R.~U.~Abbasi {\it et al.},
  %``Study of Ultra-High Energy Cosmic Ray composition using Telescope Array’s Middle Drum detector and surface array in hybrid mode,''
  Astropart.\ Phys.\  {\bf 64}, 49 (2015).
  %%CITATION = doi:10.1016/j.astropartphys.2014.11.004;%%


\bibitem{npdf} 
  K.~Kovarik {\it et al.},
  %``nCTEQ15 - Global analysis of nuclear parton distributions with uncertainties in the CTEQ framework,''
  Phys.\ Rev.\ D {\bf 93}, 085037 (2016)
  [arXiv:1509.00792 [hep-ph]].
  %%CITATION = doi:10.1103/PhysRevD.93.085037;%%
    
\end{thebibliography}
\end{document}